%% file: main.tex
\documentclass[12pt, article, a4paper, amsfonts, amssymb, amsmath, showkeys, nofootinbib, superscriptaddress]{revtex4-2}
\usepackage[english]{babel}
\usepackage[utf8]{inputenc}
\usepackage{algorithm}
\usepackage{algpseudocode}
\usepackage[colorinlistoftodos, color=green!40, prependcaption]{todonotes}
\input{preamble}
\usepackage[pdftex, pdftitle={Article}, pdfauthor={Author}]{hyperref} 
\begin{document}
\title{Genetically Engineered Quantum Circuits for Financial Market Indicators}

\author{Floyd M. Creevey}
\email{floyd.creevey@unimelb.edu.au}
\affiliation{School of Physics, University of Melbourne, VIC, Parkville, 3010, Australia.}
\author{Lloyd C. L. Hollenberg}
\email{lloydch@unimelb.edu.au}
\affiliation{School of Physics, University of Melbourne, VIC, Parkville, 3010, Australia.}

\date{\today} 

\begin{abstract}
    Quantum computing holds immense potential for transforming financial analysis and decision-making. Realising this potential necessitates the efficient encoding and processing of financial data on quantum computers. In this study, we propose using the GASP (Genetic Algorithm for State Preparation) framework to optimise the encoding of stock price data into quantum states and show it can enhance both the fidelity and efficiency of the encoding process. We demonstrate the efficacy of our approach by encoding stock price data onto both a simulated and real quantum computer to calculate the Singular Value Decomposition (SVD) entropy. Our results show improvements in fidelity and the potential for more precise financial analysis. This research provides insights into the applicability of GASP for the efficient encoding of real-world data, specifically stock price data, which is crucial for quantum advantage on noisy intermediate-scale quantum (NISQ) era quantum computers. 
\end{abstract}


\maketitle

  \section{Introduction} \label{sec:fin_introduction}

    The financial sector is always looking for innovative tools and methodologies to address the challenges of risk assessment, portfolio optimisation, and decision-making. Quantum computing has emerged as a promising technology that could impact a range of financial analysis problems. However, the transformative power of quantum computing in the finance industry depends on its ability to efficiently encode and process vast and complex financial datasets. This paper presents the theoretical framework and demonstrates the practical effectiveness of using the genetic algorithm for state preparation (GASP) \cite{creevey_gasp_2023} for encoding stock price data onto quantum states over multiple qubits. To extract meaningful insights from this quantum-encoded data, Quantum Singular Value Decomposition (QSVD) \cite{bravo-prieto_quantum_2020} is used to compute the Singular Value Decomposition (SVD) entropy of the correlation matrix. This is an important indicator in various financial analyses \cite{caraiani_predictive_2014}. The results show improvements in fidelity, computational efficiency, and the potential for more accurate financial predictions. By showing how GASP is useful for financial data encoding, valuable insights into the future of quantum finance are presented with implications spanning various financial applications, from risk management and asset allocation to market forecasting. This process requires precision and strategies that enhance the fidelity and efficiency of encoding, hence efficient and accurate generation of arbitrary quantum states with low depth and gate count is crucial for the future applications of quantum computing. Previous works have presented methods for state preparation that produce exact states \cite{niemann_logic_2016, shende_synthesis_2006, plesch_quantum-state_2011, schlimgen_quantum_2021, schuld_quantum_2019}, however, the circuits produced by these methods generally have a depth that is infeasible to execute on real hardware. Other techniques reduce the amount of depth required by introducing ancillary qubits \cite{rosenthal_query_2021, zhao_state_2019}, another parameter that is limited on current hardware. However, exact state synthesis is not always a requirement for quantum algorithms, as demonstrated in \cite{nakaji_approximate_2022}, which gives a further advantage to approximate state preparation techniques that have been developed such as \cite{melnikov_quantum_2023, araujo_low-rank_2021}. Another prominent method utilises general adversarial networks \cite{zoufal_quantum_2019}.
    

    This work builds on work presented in \cite{nakaji_approximate_2022}, which proposed an approximate amplitude encoding technique for financial market indicators and presented an alternate method for data loading. The current work follows a similar approach and demonstrates that GASP can generate circuits with sufficient accuracy to encode the stock market data required for computing the SVD entropy of stock price dynamics. Moreover, the circuits produced by GASP have a low gate depth, making them feasible to run on actual hardware. This enables analysis of the QSVD algorithm on real-world hardware. The results presented will use the methods from \cite{nakaji_approximate_2022} and \cite{zoufal_quantum_2019} for comparison.

    The rest of this paper is organised as follows; Section \ref{sec:stock_data} will describe how the stock market data statevector to be loaded is generated. Section \ref{sec:QSVD} will describe the variational quantum singular value decomposition (VQSVD) in detail, and Section \ref{sec:fin_data_loading} will discuss current methods for data loading in VQSVD. Section \ref{sec:fin_methods} will explain the methodology of using GASP for data loading in VQSVD, and Section \ref{sec:fin_conclusion} conclusions and potential future work.

  \section{Stock Price Data, Returns, and Correlation Matrices}\label{sec:stock_data}

\begin{figure*}
      \centering
        \includegraphics[width=\textwidth]{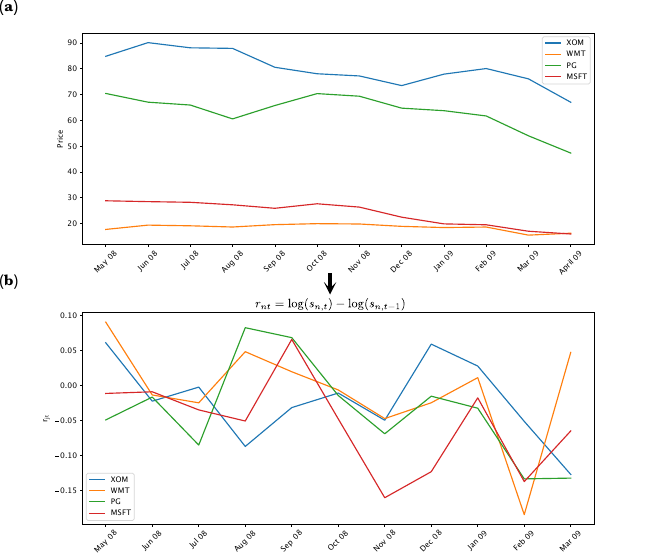}
        \caption[Process to Generate Logarithmic Rates of Return]{(a) Shows monthly stock opening price data for XOM, WMT, PG, and MSFT, for the period between March 2008 to March 2009, retrieved from Yahoo Finance. (b) Shows the calculated logarithmic rates of return $r_{nt} = \ln(s_{n, t}) - \ln(s_{n, t - 1})$ for each $n$ stock at time $t$.}
        \label{fig:cor_flow}
    \end{figure*}

\begin{figure*}
      \centering
        \includegraphics[width=0.9\textwidth]{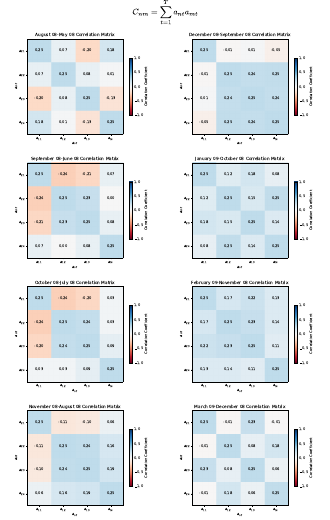}
        \caption[Process to Generate Correlation Matrices]{Shows the generated correlation matrices for XOM, WMT, PG, and MSFT, for the period between March 2008 to March 2009, retrieved from Yahoo Finance, $T=5$, $C_{nm} = \sum_{t=1}^Ta_{nt}a_{mt}$.}
        \label{fig:cor_flow_mats}
    \end{figure*}

    For some number of stocks, $n=1, 2, ..., N_s$, over some time period, $t=1, 2, ..., T$, the logarithmic rate of return can be written,
    \begin{equation}
      r_{nt} = \ln(s_{n, t}) - \ln(s_{n, t - 1}),
    \end{equation}
    where $s_{n, t}$  is the price of the $n$th stock at time $t$ \cite{caraiani_predictive_2014}. This logarithmic rate of return can be calculated between stock time points, meaning there are $T$ logarithmic rates of return for each of the $N$ stocks. The logarithmic rates of return can then be used to calculate the correlation matrix,
    \begin{equation}\label{eqn:correlation}
      C_{nm} = \sum_{t=1}^Ta_{nt}a_{mt},
    \end{equation}
    where,
    \begin{equation}\label{eqn:corr_mat_as}
      a_{nt} = \frac{r_{nt} - \langle r_n\rangle}{\sigma_n\sqrt{N_sT}},
    \end{equation}
    and $\langle r_n\rangle$ and $\sigma_n$ are the average and standard deviation, defined as,
    \begin{equation}
      \langle r_n\rangle = \frac{1}{T}\sum_{t=1}^Tr_{nt},\;\sigma_n^{2} = \frac{1}{T}\sum_{t=1}^T(r_{nt}-\langle r_n\rangle)^2.
    \end{equation}
    In Figure \ref{fig:cor_flow}(a) we show the time series data for stocks for Exxon Mobile (XOM), Walmart (WMT), Procter and Gamble (PG), and Microsoft (MSFT), and the corresponding log rates of return in Figure \ref{fig:cor_flow}(b). In Figure \ref{fig:cor_flow_mats} we give the associated correlation matrices for these time series stock data. 

\begin{figure*}
      \centering
        \includegraphics[width=\textwidth]{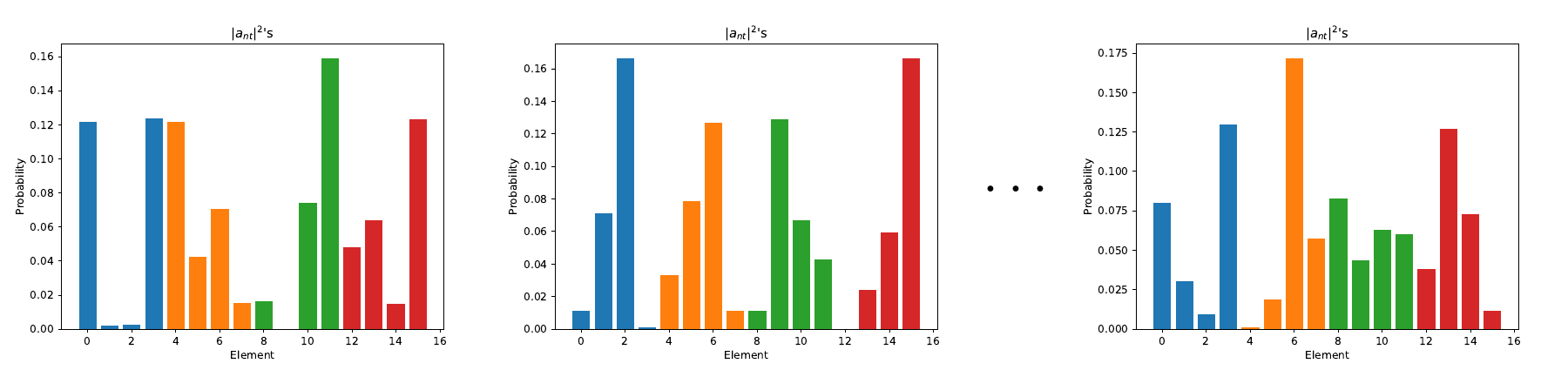}
        \caption[Finance Data Statevectors]{Shows the generated statevectors from the correlation matrices of XOM, WMT, PG, and MSFT, for the period between March 2008 to March 2009, retrieved from Yahoo Finance, $T=5$, $C_{nm} = \sum_{t=1}^Ta_{nt}a_{mt}$.}
        \label{fig:fin_data_encoding}
    \end{figure*}

    The correlation matrix, $C$, is positive semidefinite, so has nonnegative eigenvalues and satisfies,
    \begin{equation}\label{eqn:normalisation}
      \text{Tr}(C) = \sum_{n=1}^{N_s}\sum_{t=1}^{T}a_{nt}^2=1,
    \end{equation}
    so for the positive eigenvalues $\lambda_1$, $\lambda_2$, \ldots, $\lambda_{K}$, which satisfy $\sum_{k=1}^K\lambda_k=1$, the SVD entropy is defined,
    \begin{equation}
      S = -\sum_{k=1}^K\lambda_k\ln(\lambda_k).
    \end{equation}
    These $a_{nt}$ values can be loaded into  a $|\rm data\rangle$ state vector for use on a quantum computer \cite{nakaji_approximate_2022},
    \begin{equation}\label{eqn:data_state}
      |{\rm data}\rangle = \sum_{n=1}^{N_s}\sum_{t=1}^Ta_{nt}|n\rangle_{\rm{stock}}|t\rangle_{\rm{time}},
    \end{equation}
    where $\{|n\rangle\}_{j=1, 2, \ldots, N_s}$ is the basis set in the stock index Hilbert space $\mathcal{H}_\text{stock}$ and \\$\{|t\rangle\}_{t=1, 2, \ldots, T}$ is the basis set in the time index Hilbert space $\mathcal{H}_\text{time}$. Figure \ref{fig:fin_data_encoding} displays the resultant statevector encoding of the generated correlation matrices. Note the $|\rm{data}\rangle$ state defined by Equation \ref{eqn:data_state} is normalised as a result of Equation \ref{eqn:normalisation}. 
    
    Taking the partial trace over $\mathcal{H}_\text{time}$ allows the extraction of $\rho_\text{stock}$,
    \begin{align}
      \rho_\text{stock} &= \text{Tr}_{\mathcal{H}_{\text{time}}}(|\text{data}\rangle\langle\text{data}|) \\&= \sum_{nm}C_{nm}|n\rangle_\text{stock}\langle m|_\text{stock},
    \end{align}
    where $C_{nm}$ is the $(n, m)$ element of the correlation matrix given in Equation \ref{eqn:correlation}, allowing the state  $\rho_\text{stock}(C)$ to be calculated of a quantum computer. Diagonalising $\rho_\text{stock}$ allows the entropy to be calculated via VQSVD described in Section \ref{sec:QSVD}.

    To understand how VQSVD is used to diagonalise $\rho_\text{stock}$, the Schmidt decomposition must be described. The Schmidt decomposition is used to represent a bipartite quantum state as a sum of products of two orthogonal states. Consider a composite quantum system composed of two subsystems, denoted as $A$ and $B$. The Schmidt decomposition of a joint state $|\psi\rangle_{AB}$ is given by:
    \begin{equation}
      |\psi\rangle_{AB} = \sum_{i=1}^{N} \sqrt{\lambda_i} |a_i\rangle_A \otimes |b_i\rangle_B,
    \end{equation}
    where $\lambda_i$ are non-negative singular values, and $|a_i\rangle_A$ and $|b_i\rangle_B$ are orthonormal states in the respective subsystems. The relevance of this algorithm  is the Schmidt decomposition of $|\text{data}\rangle$ is equivalent to the diagonalisation of $\rho_\text{stock}(C)$,
    \begin{equation}
      |\text{data}\rangle = \sum_{j=1}^Jc_j|v_j\rangle_{\text{stock}}|v_j'\rangle_{\text{time}},
    \end{equation}
    where $\{c_j\}^J_{j=1}$  are the Schmidt coefficients, and $\{|v_j\rangle_\text{stock}\}^J_{j=1}$  and $\{|v'_j\rangle_\text{time}\}^J_{j=1}$  are sets of orthogonal states, which,  in general are not the computational basis. In this representation, $\rho_\text{stock}$ is,
    \begin{equation}
      \begin{aligned}
        \rho_\text{stock} &= \text{Tr}_{\mathcal{H}_\text{time}}(|\text{data}\rangle\langle\text{data}|)\\
        &= \sum^J_{j=1}|c_j|^2|v_j\rangle_\text{stock}\langle v_j|_\text{stock}.
      \end{aligned}
    \end{equation}
    With this exact diagonalisation of $\rho_\text{stock} = C$ it can be seen $\lambda_m = |c_m|^2$  for all $m=1, 2, \dots, J = K$  allowing the SVD entropy to be calculated as,
    \begin{equation}
      S = -\sum_{j=1}^J|c_j|^2\ln(|c_j|^2),
    \end{equation}
    which coincides with the Von Neumann entropy of $\rho_{\rm{stock}}$,
    \begin{equation}
      -\rm{Tr}(\rho_{\rm{stock}}\ln(\rho_{\rm{stock}})).
    \end{equation}

  \section{Variational Quantum Singular Value Decomposition (VQSVD)}\label{sec:QSVD}

    Before describing VQSVD, first the singular value decomposition (SVD) must be described. SVD decomposes any given $m\times n$ complex matrix, $A$, as,
    \begin{equation}
      A = U\Sigma V^\dag,
    \end{equation}
    where $U$ is an $m\times m$ complex unitary matrix, $\Sigma$ is an $m\times n$ rectangular diagonal matrix with non-negative real numbers on the diagonal, $V$ is an $n\times n$ complex unitary matrix, and $V^\dag$ is the Hermitian conjugate of $V$. If $A$ is real, $U$ and $V$ can be guaranteed to be orthogonal matrices; in this case the SVD is often,
    \begin{equation}
      A = U\Sigma V^T.
    \end{equation}

    VQSVD is a quantum algorithmic approach that combines quantum circuits with classical optimisation to find an approximate singular value decomposition of a given matrix \cite{bravo-prieto_quantum_2020}. Given a matrix $M$, VQSVD seeks matrices $U(\vec{\theta})$, $\Sigma$, and $V(\vec{\theta}')$ such that:
    \begin{equation}
      M \approx U(\vec{\theta}) \Sigma V(\vec{\theta'})^\dagger,
    \end{equation}
    where $U(\vec{\theta})$ and $V(\vec{\theta'})$ are unitary matrices, and $\Sigma$ is a diagonal matrix with singular values on the diagonal. 
    
    VQSVD is used to transform the Schmidt basis $\{|v_j\rangle_\text{stock}\}^J_{j=1}$  and $\{|v'_j\rangle_\text{time}\}^J_{j=1}$  to the computational basis, so that the values of $\{c_j\}^J_{j=1}$  can be extracted. To realise this, after preparing the quantum state $|\text{data}\rangle$, a variational quantum circuit that approximates the unitary matrices $U(\vec{\theta})$ and $V(\vec{\theta'})$ are applied such that ideally,
    \begin{equation}
      U(\vec{\theta})\otimes V(\vec{\theta'})|\widetilde{\text{data}}\rangle = \sum^J_{j=1}c_j|\tilde{j}\rangle_\text{stock}|\tilde{j}\rangle_\text{time},
    \end{equation}
    with $\{|\tilde{j}\rangle_\text{stock}\}^J_{j=1}$  and $\{|\tilde{j}\rangle_\text{time}\}^J_{j=1}$  being subsets of the computational basis state satisfying $\langle\tilde{j}|\tilde{k}\rangle = \delta_{j, k}$. This identifies the Schmidt basis as $|v_j\rangle_\text{stock} = U(\theta)|\tilde{j}\rangle_\text{stock}$ and $|v'_j\rangle_\text{time} = V(\theta')^\dagger|\tilde{j}\rangle_\text{time}$. Classical optimisation techniques are then used to update the circuit parameters to minimise the difference between $U(\vec{\theta})\otimes V(\vec{\theta'})|\text{data}\rangle$ and \newline$\sum^J_{j=1}c_j|\tilde{j}\rangle_\text{stock}|\tilde{j}\rangle_\text{time}$. This is done by summing the Hamming distances between the stock bit sequences and time bit sequences, obtained as a result of computational-basis measurements on $\mathcal{H}_\text{stock}$ and $\mathcal{H}_\text{time}$, which can be expressed as the cost function, 
    \begin{equation}
      \mathcal{L}_{\rm SVD}(\theta, \theta') = \sum_{q=1}^{n_s}\frac{1 - \langle\sigma_z^q\sigma_z^{q+n_s}\rangle}{2},
    \end{equation}
    with $n_s = \log(N_s)$, $\langle\cdot\rangle$ being the expectation over $U(\vec{\theta})\otimes V(\vec{\theta'})|{\rm data}\rangle$, and $\sigma_z^q$ being the Pauli Z operator acting on the $q$th qubit. When $\mathcal{L}_{\rm SVD}(\theta, \theta') = 0$, $U(\vec{\theta})\otimes V(\vec{\theta'})|\text{data}\rangle$ is equal to $\sum^J_{j=1}c_j|\tilde{j}\rangle_\text{stock}|\tilde{j}\rangle_\text{time}$, meaning that minimising $\mathcal{L}_{\rm SVD}(\theta, \theta')$ creates better approximations of the Schmidt decomposed state.

  \section{Current Methods for Data Loading in VQSVD}\label{sec:fin_data_loading}

    Before describing GASP for data loading in VQSVD, it is important to describe the methods used for comparison, i.e the quantum generative adversarial networks (qGAN) method \cite{zoufal_quantum_2019}, and the approximate amplitude encoding (AAE) method \cite{nakaji_approximate_2022}.

    \textbf{\textit{qGAN Method}}

      qGAN is a technique that combines quantum and classical computing to approximately load quantum states. It leverages quantum generative adversarial networks to enable the effective learning and encoding of various probability distributions, which are defined by data samples, into quantum states. The approach employs a quantum channel, such as a variational quantum circuit, or quantum generator, alongside a classical neural network, allowing the system to capture the underlying probability distribution of the data samples and encode it into a quantum state. The quantum generator is trained to transform an $n$-qubit input state, $|\psi_{\rm{in}}\rangle$, to an $n$-qubit output state,

      \begin{equation}
        G_\theta|\psi_{\rm{in}}\rangle = |g_\theta\rangle = \sum_{j=0}^{2^n-1}\sqrt{p_\theta^j}|j\rangle,
      \end{equation}

      where $p_\theta^j$ describes the resulting occurrence probabilities of the basis states $|j\rangle$. The quantum generator, implemented as a parameterised quantum circuit, consists of alternating layers of $R_y$ gates and controlled-Z, $CZ$, gates.

    \textbf{\textit{AAE Method}}

      AAE expands on the qGAN approach by loading all parts of a real-valued data vector, including both magnitude and sign, into the amplitude of a quantum state, whereas qGAN only captures the magnitudes. This is accomplished through variational training of a shallow parameterised quantum circuit, utilising two types of measurements: standard computational-basis measurements and Hadamard-transformed basis measurements, the latter specifically addressing the signs of the data components. The variational algorithm adjusts the circuit parameters to minimise the sum of two costs associated with these measurement bases, both of which are derived from the efficiently computable maximum mean discrepancy (MMD). For some $|j\rangle = |j_1j_2\ldots j_n\rangle$ where $j_k$ is the state of the $k$th qubit in the computational basis and $j=\sum_{k=1}^n2^{n-k}j_k$, the MMD is a measure of the difference between two probability distributions, the model distribution, $q_\theta(j)$, and the target distribution, $p(j)$. This allows the parameterised quantum circuit to be minimised with a cost function $\mathcal{L}_{\rm{MMD}}(q_\theta, p)$, defined as,
      \begin{equation}
        \begin{aligned}
          \mathcal{L}_{\rm{MMD}}(q_\theta, p)&\equiv\gamma_{\rm{MMD}}(q_\theta, p)^2 \\
          \gamma_{\rm{MMD}}(q_\theta, p) &= \left|\sum_{j=0}^{2^n-1}q_\theta(j)\Phi(j)-\sum_{j=0}^{2^n-1}p(j)\Phi(j)\right|,
        \end{aligned}
      \end{equation}

      where $\Phi(j)$ is a function that maps $j$ to a feature space. The parameterised quantum circuits consist of alternating layers of $R_y$ gates and $CNOT$ gates connecting adjacent qubits.

  \section{GASP for Data Loading in VQSVD}\label{sec:fin_methods}

\begin{figure*}
      \centering
        \includegraphics[width=\textwidth]{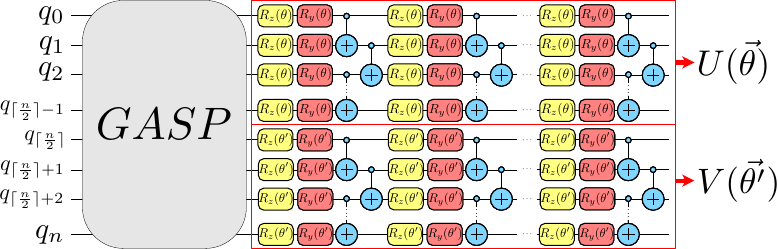}
        \caption[QSVD Experimental Circuit Structure]{QSVD Experimental circuit structure. The GASP block represents the state encoding circuit produced by GASP. The QSVD section of the circuit is separated into two pieces, the circuit for $U(\vec{\theta})$, and the circuits for $V(\vec{\theta'})$. These circuits are each composed of an $R_z(\theta)$ and $R_y(\theta)$ gate on each qubit, followed by $\rm{CNOT}$ gates to create linear entanglement over the qubits. The $\theta$'s are all initialised as random values. These layers can be repeated to improve the QSVD accuracy.}
        \label{fig:QSVD_circuit}
    \end{figure*}

    The parameterised quantum circuit is composed of the GASP generated $|{\rm data}\rangle$ unitary circuit, as well as the $U(\vec{\theta})$ and $V(\vec{\theta'})$ circuits of the QSVD, each composed of parameterised single-qubit rotational gates, $R_z(\theta_i)$ and $R_y(\theta_i)$, and CNOT gates that connect adjacent qubits. The structure of the full circuit is shown in Figure \ref{fig:QSVD_circuit}.

    For demonstration, monthly stock opening price data for XOM, WMT, PG, and MSFT was retrieved from Yahoo Finance for the period between March 2008 to March 2009, displayed in Table \ref{table:stock_data} and the results of the logarithmic rates of return calculations are shown in Figure \ref{fig:cor_flow}(b). 


    \begin{table}[h]
  \small
  \centering
  \caption{\label{table:stock_data_inverted}Stock Prices for Selected Symbols (Apr 08 - Mar 09)}
  \begin{tabular}{@{}l|llllllllllll}
      \hline
      \hline
      Symbol & Apr 08 & May 08 & Jun 08 & Jul 08 & Aug 08 & Sep 08 & Oct 08 & Nov 08 & Dec 08 & Jan 09 & Feb 09 & Mar 09 \\
      \hline
      XOM  & 84.80 & 90.10 & 88.09 & 87.87 & 80.55 & 78.04 & 77.19 & 73.45 & 77.89 & 80.06 & 76.06 & 67.00 \\
      WMT  & 53.19 & 58.20 & 57.41 & 56.00 & 58.75 & 59.90 & 59.51 & 56.76 & 55.37 & 55.98 & 46.57 & 48.81 \\
      PG   & 70.41 & 67.03 & 65.92 & 60.55 & 65.73 & 70.35 & 69.34 & 64.72 & 63.73 & 61.69 & 54.00 & 47.32 \\
      MSFT & 28.83 & 28.50 & 28.24 & 27.27 & 25.92 & 27.67 & 26.38 & 22.48 & 19.88 & 19.53 & 17.03 & 15.96 \\
      \hline
      \hline
  \end{tabular}
\end{table}

\begin{figure*}
      \centering
        \includegraphics[width=\textwidth]{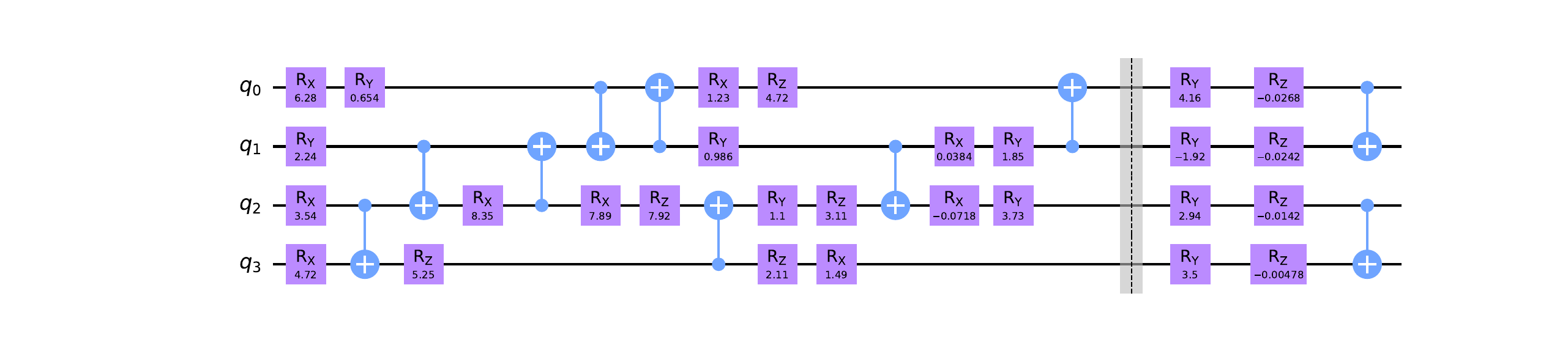}
        \caption[Example QSVD Circuit with GASP Statevector Initialisation]{Example of an experimental VQSVD circuit with GASP as the state vector initialisation circuit generation technique. The GASP-generated circuit is displayed before the barrier (dashed vertical black line), $U(\vec{\theta})$ and $V(\vec{\theta'})$ composing the QSVD circuit are displayed after the barrier. Circuits generated by GASP will almost always be unique, because of the sporadic nature of genetic algorithms, hence why this is just one example of the generated experimental circuits.}
        \label{fig:opt_QSVD}
    \end{figure*}

    This data can be encoded on $4$ qubits. These logarithmic rates of return were then used to calculate the SVD entropy over a period of 5 months. As there were four stocks and five months, $N=4$, and $T=4$. As a result, the $a_{jt}$ values formed a four-by-four matrix, allowing loading into a four qubit state vector as,
    \begin{equation}
      |{\rm data}\rangle = \sum_{n=1}^{4}\sum_{t=1}^{4}a_{nt}|n\rangle|t\rangle.
    \end{equation}
    With $N=4$ and $T=4$, $n_s = \log(4) = 2$, and $t_s = \log(4) = 2$. This $|\rm{data}\rangle$ was the target vector given to GASP with which to generate an initialisation circuit. With $4$ stocks and $4$ time steps, the cost function was of the form,
    \begin{equation}
      \mathcal{L}_{\rm SVD}(\theta, \theta') = \sum_{q=1}^{2}\frac{1 - \langle\sigma_z^q\sigma_z^{q+2}\rangle}{2}.
    \end{equation}
    The VQSVD circuits $U(\vec{\theta})$ and $V(\vec{\theta'})$ used each had one layer, and were optimised with SPASA. This was repeated for each possible five-month section, at GASP target fidelities of $70\%$, $75\%$, $80\%$, $85\%$, $90\%$, $95\%$, and $99\%$. The circuit for the section of data from April to August 2008 is shown in Figure \ref{fig:opt_QSVD} as an example. The results generated using GASP for the state preparation were compared against the qGAN and AAE methods.
  
\begin{figure*}
      \centering
        \includegraphics[width=\textwidth]{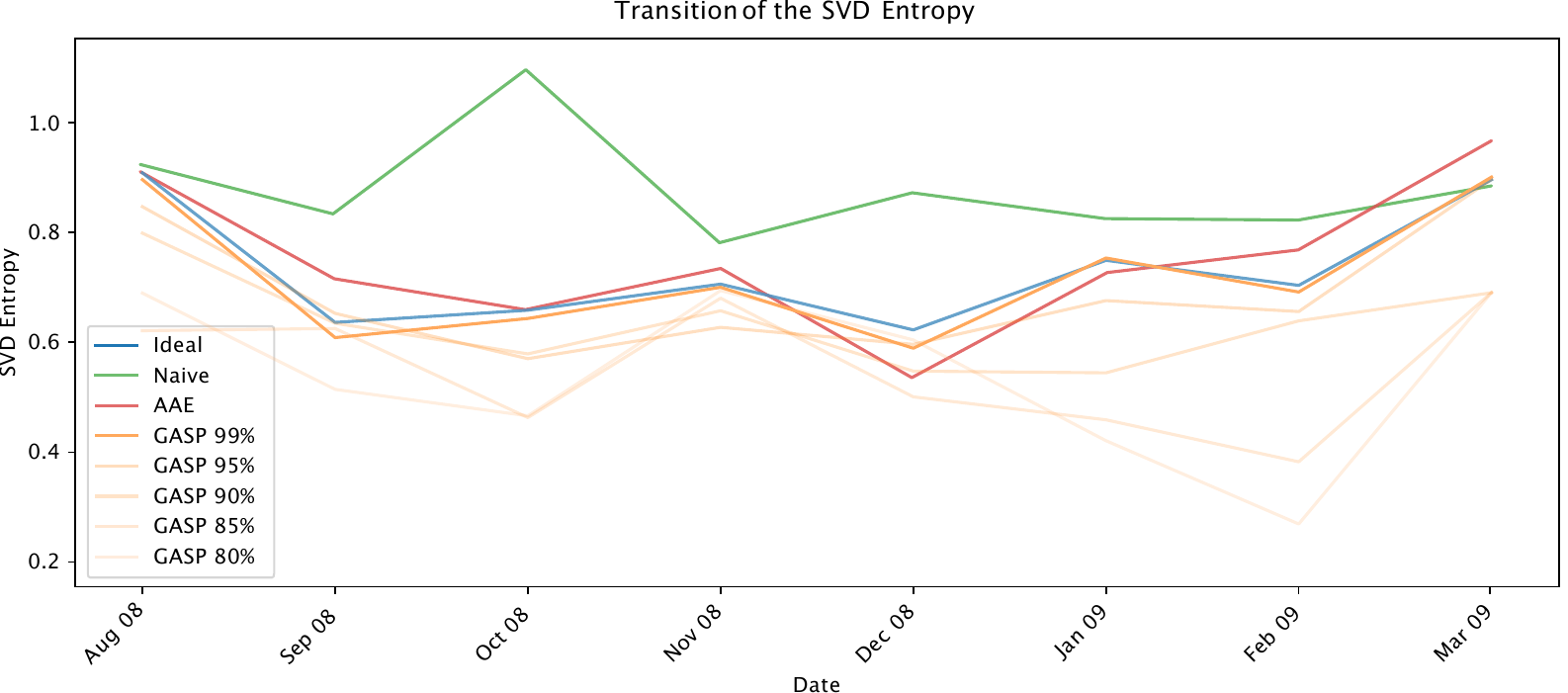}
        \caption[Comparison of SVD Entropy]{Comparison of SVD entropy between each section for each method, e.g `Aug 08' shows the calculated SVD entropy for the XOM, WMT, PG, MSFT stocks over the Apr 08, May 08, June 08, July 08 time period. The ideal SVD entropies (calculated by diagonalising the correlation matrix), qGAN, AAE, and GASP, are shown in blue, green, red, and orange, respectively. Higher target fidelity GASP results are shown with decreasing transparency.}
        \label{fig:SVD_ent}
    \end{figure*}

    Analysing the results, it can be seen that the higher fidelity GASP circuits produced more accurate singular values than qGAN or AAE, as shown in Figure \ref{fig:SVD_ent}, which lead to improvements in the reconstruction of the correlation matrix. 
    
\begin{figure*}
      \centering
        \includegraphics[width=\textwidth]{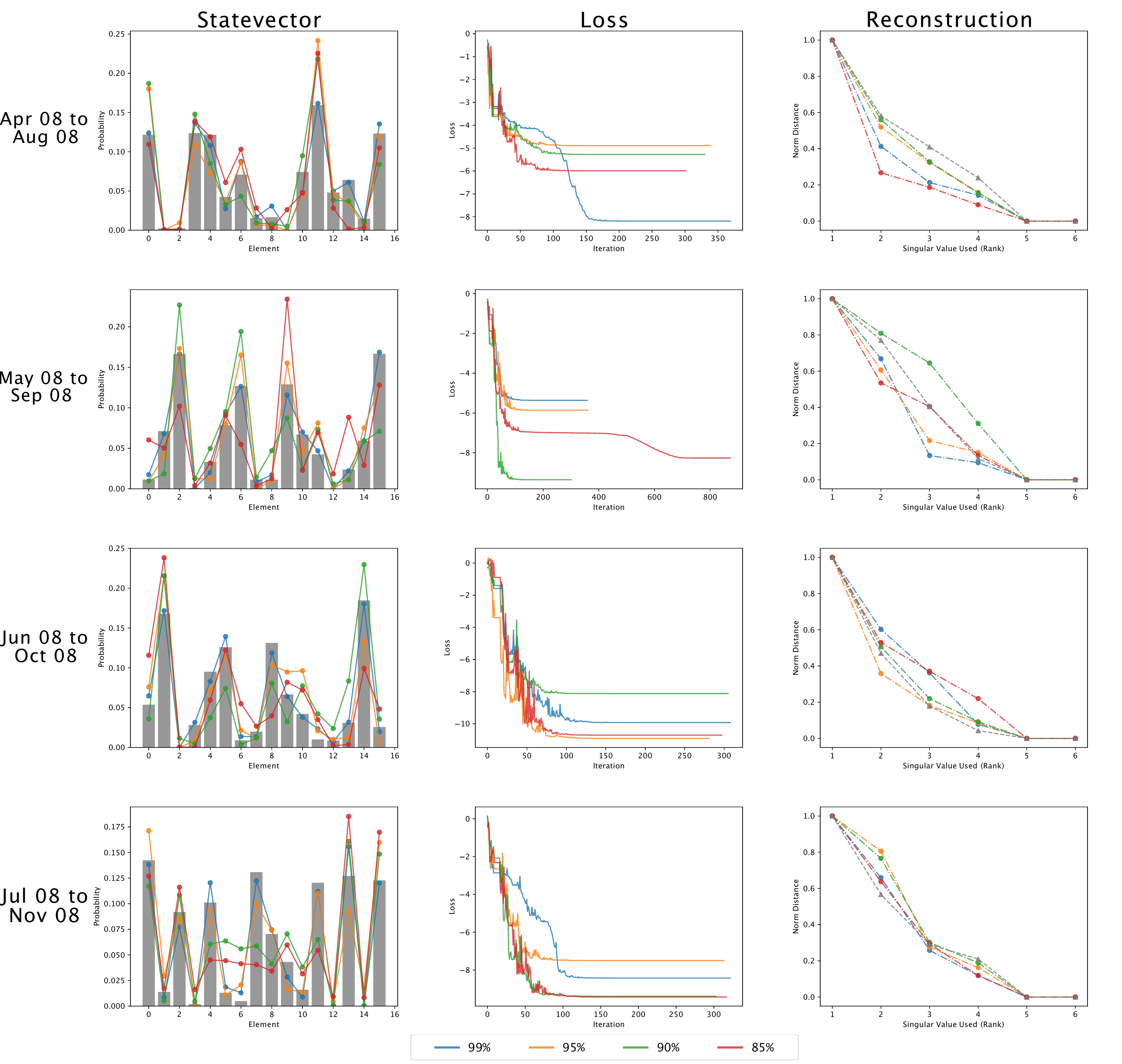}
        \caption[Example QSVD Results with GASP Statevector Initialisation]{Example GASP results for the XOM, WMT, PG, MSFT stocks over the periods of Apr 08 to Aug 08, May 08 to Sep 08, Jun 08 to Oct 08, and Jul 08 to Nov 08. The first column shows the target $|\rm{data}\rangle$ state vector with grey bars, and the achieved state vectors. The second column shows the loss plots. The third column shows the Frobenius-norm error of the reconstructed correlation matrix. $99\%$, $95\%$, $90\%$, and $85\%$ fidelities are shown in blue, orange, green, and red, respectively.}
        \label{fig:QSVD_res}
    \end{figure*}

    The data statevectors generated with GASP for the periods of Apr 08 to Aug 08, May 08 to Sep 08, Jun 08 to Oct 08, and Jul 08 to Nov 08 can be seen in the first column of Figure \ref{fig:QSVD_res}. It can be seen that as the target fidelity increases, the generated statevector more accurately represents the target statevector, $|\rm{data}\rangle$. The second column of Figure \ref{fig:QSVD_res} shows the loss plots for the optimisation process to produce the singular values. The third column shows the Frobenius-norm reconstruction error for each target fidelity statevector in reconstructing the singular values.
    
\begin{figure*}
      \centering
        \includegraphics[width=\textwidth]{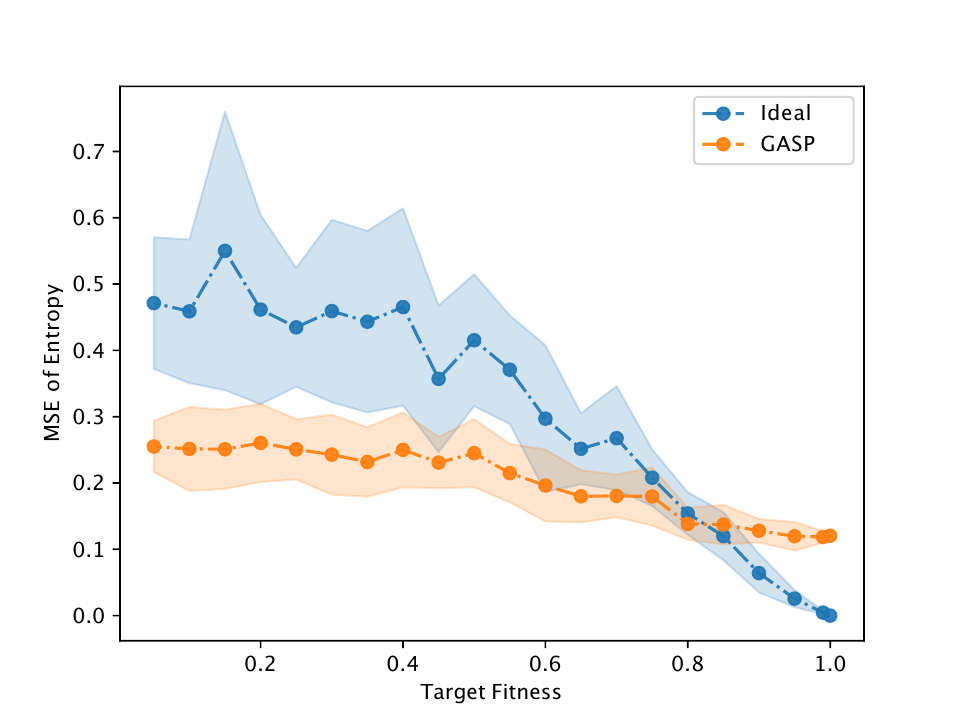}
        \caption[Mean Square Error Results for SVD Entropy]{Mean square error results for SVD entropy calculated over target fitnesses of ($0.05\%$ - $0.99\%$) in increments of $5\%$ for ideal (diagonalised correlation matrix) and GASP  (QSVD) SVD entropy (calculated by diagonalising the correlation matrix).}
        \label{fig:MSE}
    \end{figure*}

    When viewing the mean squared error of the SVD entropy for each of the $4$ qubit GASP circuits with the ideal SVD entropy (calculated by diagonalising the correlation matrix), it becomes apparent that there are minimal improvements in the accuracy of the calculated SVD entropy above $90\%$. As the increase in time taken, $\rm{CNOT}$ gates, and total gates, for GASP to generate a $99\%$ fidelity circuits compared to $\approx90\%$ circuits is far greater, results that imply lower fidelity circuits performing similarly are promising. This trend can be seen in the results plotted in Figure \ref{fig:MSE}. 

  \section{Conclusion}\label{sec:fin_conclusion}

    The primary focus of this investigation was to determine how well GASP performs in generating approximate state preparations for financial quantum algorithms, specifically QSVD for calculating SVD entropy. The findings demonstrate that it is feasible to sacrifice perfect state preparation for approximate state preparation without significantly compromising the accuracy of the calculated SVD entropy. The main result highlights that moving from perfect state preparation to approximate state preparation, the mean squared error in the calculated SVD entropy gradually increases, but the rate of increase diminishes. This observation suggests that the diminishing returns associated with achieving perfect state preparation may not be a worthwhile pursuit in practical financial applications. Instead, adopting approximate state preparation strategies, which are computationally efficient, can offer an appealing trade-off between accuracy and resource utilisation. The implications of this research are substantial for the field of quantum finance and beyond. By leveraging GASP-generated circuits, efficient and practical financial algorithms can be implemented on NISQ-era quantum computers. This approach can enhance risk management, asset allocation, and market forecasting in the financial sector, making it a valuable tool for investors, analysts, and financial institutions. This study demonstrates the potential of using GASP in financial quantum algorithms. The balance between accuracy and efficiency is a critical consideration, and the results indicate that approximations in state preparation can yield accurate financial predictions. As the field of quantum finance continues to evolve, these findings open exciting avenues for further research and practical implementation, ultimately contributing to the development of more robust and efficient financial market indicators in the quantum era.
    
    \bibliography{QSVD.bib}

\end{document}

%% file: preamble.tex
\usepackage{amsthm}
\usepackage{mathtools}
\usepackage{physics}
\usepackage{xcolor}
\usepackage{graphicx}
\usepackage[margin=0.75in]{geometry}
\usepackage{adjustbox}
\usepackage{placeins}
\usepackage[T1]{fontenc}
\usepackage{lipsum}
\usepackage{csquotes}

%% file: QSVD.bib
@inproceedings{niemann_logic_2016,
	title = {Logic {Synthesis} for {Quantum} {State} {Generation}},
	doi = {10.1109/ISMVL.2016.30},
	booktitle = {2016 {IEEE} 46th {International} {Symposium} on {Multiple}-{Valued} {Logic} ({ISMVL})},
	author = {Niemann, Philipp and Datta, Rhitam and Wille, Robert},
	month = may,
	year = {2016},
	note = {ISSN: 2378-2226},
	pages = {247--252},
}

@article{shende_synthesis_2006,
	title = {Synthesis of {Quantum} {Logic} {Circuits}},
	volume = {25},
	issn = {0278-0070, 1937-4151},
	url = {http://arxiv.org/abs/quant-ph/0406176},
	doi = {10.1109/TCAD.2005.855930},
	number = {6},
	urldate = {2021-10-07},
	journal = {IEEE Transactions on Computer-Aided Design of Integrated Circuits and Systems},
	author = {Shende, Vivek V. and Bullock, Stephen S. and Markov, Igor L.},
	month = jun,
	year = {2006},
	note = {arXiv: quant-ph/0406176},
	pages = {1000--1010},
}

@article{nakaji_approximate_2022,
	title = {Approximate amplitude encoding in shallow parameterized quantum circuits and its application to financial market indicators},
	volume = {4},
	issn = {2643-1564},
	url = {https://link.aps.org/doi/10.1103/PhysRevResearch.4.023136},
	doi = {10.1103/PhysRevResearch.4.023136},
	
	number = {2},
	urldate = {2023-05-15},
	journal = {Physical Review Research},
	author = {Nakaji, Kouhei and Uno, Shumpei and Suzuki, Yohichi and Raymond, Rudy and Onodera, Tamiya and Tanaka, Tomoki and Tezuka, Hiroyuki and Mitsuda, Naoki and Yamamoto, Naoki},
	month = may,
	year = {2022},
	pages = {023136},
}

@article{plesch_quantum-state_2011,
	title = {Quantum-state preparation with universal gate decompositions},
	volume = {83},
	issn = {1050-2947, 1094-1622},
	url = {https://link.aps.org/doi/10.1103/PhysRevA.83.032302},
	doi = {10.1103/PhysRevA.83.032302},
	
	number = {3},
	urldate = {2023-05-15},
	journal = {Physical Review A},
	author = {Plesch, Martin and Brukner, Caslav},
	month = mar,
	year = {2011},
	pages = {032302},
}

@article{schlimgen_quantum_2021,
	title = {Quantum {Simulation} of {Open} {Quantum} {Systems} {Using} a {Unitary} {Decomposition} of {Operators}},
	volume = {127},
	url = {https://link.aps.org/doi/10.1103/PhysRevLett.127.270503},
	doi = {10.1103/PhysRevLett.127.270503},
	number = {27},
	urldate = {2023-05-15},
	journal = {Physical Review Letters},
	author = {Schlimgen, Anthony W. and Head-Marsden, Kade and Sager, LeeAnn M. and Narang, Prineha and Mazziotti, David A.},
	month = dec,
	year = {2021},
	note = {Publisher: American Physical Society},
	pages = {270503},
}

@article{creevey_gasp_2023,
	title = {{GASP}: a genetic algorithm for state preparation on quantum computers},
	volume = {13},
	copyright = {2023 The Author(s)},
	issn = {2045-2322},
	shorttitle = {{GASP}},
	url = {https://www.nature.com/articles/s41598-023-37767-w},
	doi = {10.1038/s41598-023-37767-w},
	
	number = {1},
	urldate = {2023-09-22},
	journal = {Scientific Reports},
	author = {Creevey, Floyd M. and Hill, Charles D. and Hollenberg, Lloyd C. L.},
	month = jul,
	year = {2023},
	note = {Number: 1
Publisher: Nature Publishing Group},
	pages = {11956},
}

@article{schuld_quantum_2019,
	title = {Quantum machine learning in feature {Hilbert} spaces},
	volume = {122},
	issn = {0031-9007, 1079-7114},
	url = {http://arxiv.org/abs/1803.07128},
	doi = {10.1103/PhysRevLett.122.040504},
	number = {4},
	urldate = {2023-10-03},
	journal = {Physical Review Letters},
	author = {Schuld, Maria and Killoran, Nathan},
	month = feb,
	year = {2019},
	note = {arXiv:1803.07128 [quant-ph]},
	pages = {040504},
}

@misc{zhao_state_2019,
	title = {State preparation based on quantum phase estimation},
	url = {https://arxiv.org/abs/1912.05335v1},
	
	urldate = {2023-10-16},
	journal = {arXiv.org},
	author = {Zhao, Jian and Wu, Yu-Chun and Guo, Guang-Can and Guo, Guo-Ping},
	month = dec,
	year = {2019},
}

@misc{rosenthal_query_2021,
	title = {Query and {Depth} {Upper} {Bounds} for {Quantum} {Unitaries} via {Grover} {Search}},
	url = {https://arxiv.org/abs/2111.07992v4},
	
	urldate = {2023-10-16},
	journal = {arXiv.org},
	author = {Rosenthal, Gregory},
	month = nov,
	year = {2021},
}

@article{melnikov_quantum_2023,
	title = {Quantum state preparation using tensor networks},
	volume = {8},
	issn = {2058-9565},
	url = {https://dx.doi.org/10.1088/2058-9565/acd9e7},
	doi = {10.1088/2058-9565/acd9e7},
	
	number = {3},
	urldate = {2023-10-16},
	journal = {Quantum Science and Technology},
	author = {Melnikov, Ar A. and Termanova, A. A. and Dolgov, S. V. and Neukart, F. and Perelshtein, M. R.},
	month = jun,
	year = {2023},
	note = {Publisher: IOP Publishing},
	pages = {035027},
}

@misc{araujo_low-rank_2021,
	title = {Low-rank quantum state preparation},
	url = {https://arxiv.org/abs/2111.03132v3},
	
	urldate = {2023-10-16},
	journal = {arXiv.org},
	author = {Araujo, Israel F. and Blank, Carsten and Araújo, Ismael C. S. and da Silva, Adenilton J.},
	month = nov,
	year = {2021},
	doi = {10.1109/TCAD.2023.3297972},
}

@article{bravo-prieto_quantum_2020,
	title = {Quantum singular value decomposer},
	volume = {101},
	url = {https://link.aps.org/doi/10.1103/PhysRevA.101.062310},
	doi = {10.1103/PhysRevA.101.062310},
	number = {6},
	urldate = {2023-10-16},
	journal = {Physical Review A},
	author = {Bravo-Prieto, Carlos and García-Martín, Diego and Latorre, José I.},
	month = jun,
	year = {2020},
	note = {Publisher: American Physical Society},
	pages = {062310},
}

@article{caraiani_predictive_2014,
	title = {The predictive power of singular value decomposition entropy for stock market dynamics},
	volume = {393},
	issn = {0378-4371},
	url = {https://www.sciencedirect.com/science/article/pii/S0378437113008212},
	doi = {10.1016/j.physa.2013.08.071},
	urldate = {2023-10-16},
	journal = {Physica A: Statistical Mechanics and its Applications},
	author = {Caraiani, Petre},
	month = jan,
	year = {2014},
	pages = {571--578},
}

@article{zoufal_quantum_2019,
	title = {Quantum {Generative} {Adversarial} {Networks} for learning and loading random distributions},
	volume = {5},
	copyright = {2019 The Author(s)},
	issn = {2056-6387},
	url = {https://www.nature.com/articles/s41534-019-0223-2},
	doi = {10.1038/s41534-019-0223-2},
	
	number = {1},
	urldate = {2023-10-16},
	journal = {npj Quantum Information},
	author = {Zoufal, Christa and Lucchi, Aurélien and Woerner, Stefan},
	month = nov,
	year = {2019},
	note = {Number: 1
Publisher: Nature Publishing Group},
	pages = {1--9},
}
